\documentstyle[amssymb,preprint,aps]{revtex}

\newtheorem{theorem}{Theorem}
\newtheorem{acknowledgement}[theorem]{Acknowledgement}

\begin{document}
\title{Boundary of the Set of Separable States}
\author{{\small Mingjun Shi\thanks{%
Email: shmj@ustc.edu.cn}, Jiangfeng Du\thanks{%
Email: djf@ustc.edu.cn}}}
\address{Laboratory of Quantum Communication and Quantum Computation,\\
Department of Modern Physics,\\
University of Science and Technology of China, Hefei, 230026, P.R.China}
\date{\today }
\maketitle

\begin{abstract}
The set of all separable quantum states is compact and convex. We focus on
the two-qubit quanum system and study the boundary of the set. Then we give
the criterion to determine whether a separable state is on the boundary.
Some straightforward geometrical interpretations for entanglement are based
on the concept and presented subsequently.

\end{abstract}

\section{INTRODUCTION}

Entanglement and nonlocality are some of the most emblematic concepts
embodied in quantum mechanics\cite{1,2}. Recent developments in quantum
computation\cite{3,4} and quantum information (e.g. \cite{5,6,7}) evoke
interest in the properties of quantum entanglement.

Now we have known that a state of a composite quantum system is called
entangled (or inseparable) if it cannot be represented as a convex
combination of tensor products of its subsystem states\cite{4,8}. In the
case of pure states, one has obtained a complete understanding, that is, a
pure state is entangled if and only if it violates Bell's\- inequality\cite%
{9,10,11}. But the case for mixed states is much more complicated (e.g.\cite%
{12,13}). Besides the above definition of entangled state, another
qualitative criterion proposed by Peres\cite{14} is more practicable to
determine whether a given state is inseparable, although it is sufficient
and necessary only for sone simple quantum system, such as $2\times 2$ and $%
2\times 3$ systems. On the other hand, one is more interested in the
quantitative measure of entangled states. Various entanglement measure has
been proposed. For example, for two-party system, there are following kinds
of measure: the entanglement of formation\cite{4}, the relative entropy of
entanglement\cite{15,16}, and the robustness of entanglement\cite{17}, {\it %
etc.}

Consider the relative entropy associated with entangled state $\rho _{e}$ of
two-qubit system. In order to obtain the relative entropy of entanglement of 
$\rho _{e}$, the crucial point is to find a separable state $\rho _{s}$ such
that minimizes the relative entropy $S\left( \rho _{e}\parallel \rho
_{s}\right) $ defined as 
\begin{equation}
S\left( \rho _{e}\parallel \rho _{s}\right) =tr\left\{ \rho _{e}\ln \frac{%
\rho _{e}}{\rho _{s}}\right\} .
\end{equation}%
In general case, to find $\rho _{s}$ is a laborious work.

Then recall the robustness of entanglement. It is said that given an
entangled state $\rho _{e}$ and a separable state $\rho _{s}$ there always
exists a minimal real $s\geqslant 0$ for which 
\begin{equation}
\rho \left( s\right) =\frac{1}{1+s}\left( \rho _{e}+s\rho _{s}\right)
\end{equation}%
is separable. Such minimal $s$, denoted by $R\left( \rho _{e}\parallel \rho
_{s}\right) $, is called robustness of $\rho _{e}$ relative to $\rho _{s}$.
Then the entanglement embodied in $\rho _{e}$ can be measured by the
absolute robustness of $\rho _{e}$, which is obtained by means of finding a
separable state $\rho _{s}^{\prime }$ such that $R\left( \rho _{e}\parallel
\rho _{s}^{\prime }\right) $ is minimal. We can see that both in relative
entropy and in robustness, finding a specific separable state is of essence.

Additionally, let's consider another useful tool to study entanglement, {\it %
i.e.}, Lewenstein-Sanpera (L-S) decomposition\cite{18}. Lewenstein {\it %
et.al.} have shown that any state of two-qubit system can be decomposed as 
\begin{equation}
\rho =\lambda \rho _{s}+\left( 1-\lambda \right) P_{e},\quad \lambda \in 
\left[ 0.1\right] ,
\end{equation}%
where $\rho _{s}{\cal \ }$is a separable state, and $P_{e}$ denotes a single
pure entangled projector $(P_{e}=\left| \Psi _{e}\right\rangle \left\langle
\Psi _{e}\right| )$, that is, $P_{e}$ represents a pure entangled state. It
is known that there are many different L-S decomposition with varying values
of $\lambda $ and various $P_{e}$'s. Among them is the unique optimal
decomposition, the one with the largest $\lambda $ value in Eq. (3), 
\begin{equation}
\rho =S\rho _{s}^{(opt)}+(1-S)P_{e}^{(opt)},
\end{equation}%
where 
\[
S=\max \left\{ \lambda \right\} 
\]%
is the degree of separability possessed by $\rho $. In other words, the
entanglement possessed by $\rho $ can be represented as $\left( 1-S\right)
E\left( \Psi _{e}\right) $, where $E\left( \Psi _{e}\right) $ is the
entanglement of its pure state expressed in terms of the von Neumann entropy
of the reduced density matrix of either of its subsystems:%
\begin{equation}
E\left( \Psi _{e}\right) =-tr\left( \rho _{A}\log \rho _{A}\right)
=-tr\left( \rho _{B}\log \rho _{B}\right) .
\end{equation}%
Here again, in the sense of maximizing or minimizing some quality, a
specific separable state $\rho _{s}^{(opt)}$ appears.

From the above consideration, we show that there are some specific separable
states playing an important role in determining the entanglement measure.
Then it is meaningful to determine the properties of these separable states.
A fact provokes our attention. That is, at least for two-qubit system the
set of separable states, denoted by ${\cal S}$, is compact and convex\cite%
{19}. Then boundary of ${\cal S}$ does exist. We argue in this paper that
these special separable states mentioned above are on the boundary of ${\cal %
S}$. In the following, we give the criterion to determine whether a
separable state is on the boundary. Taking this into account, we present
more clear interpretations on some concepts concerning quantum entanglement.

\section{BOUNDARY OF ${\cal S}$}

In this paper, we focus on the two-qubit quantum system. The states can be
represented in the Hilbert space ${\Bbb C}^{2}\otimes {\Bbb C}^{2}$. So we
also call it 2$\times $2 quantum system. We know that all $4\times 4$
Hermitian matrices span a 16-dimensional real Hilbert space, denoted by $%
{\cal H}$. All density matrices of states forms a subset of ${\cal H}$. We
choose Hilber-Schimdt metric, that is, for $A\in {\cal H}$, the norm is
defined as $\left\| A\right\| =\left[ tr\left( A^{+}A\right) \right] ^{\frac{%
1}{2}}=\left[ tr\left( A^{2}\right) \right] ^{\frac{1}{2}}$, and for $A,B\in 
{\cal H}$, the distance of $A$ and $B$ is%
\begin{equation}
\left\| A-B\right\| =\left[ tr\left( A-B\right) ^{2}\right] ^{\frac{1}{2}}.
\end{equation}

For two-qubit system,a given state $\rho $ is separable, i.e. $\rho \in 
{\cal S}$, if and only if the partial transposition of $\rho $ ,denoted by $%
\rho ^{T_{b}}$ or $\rho ^{T_{a}}$ $(\rho ^{T_{b}}=(\rho ^{T_{a}})^{\ast })$,
is also positive\cite{14,19}, or equivalently, the partial time-reversal of $%
\rho $, denoted by $\widetilde{\rho }$, is positive. Consider an arbitrary
separable state $\rho _{s}\in {\cal S}$. Mixing it with the maximal
separable state $\rho _{0}=\frac{1}{4}I_{4\times 4}$, we have 
\begin{equation}
\rho _{\varepsilon }=\left( 1-\varepsilon \right) \rho _{0}+\varepsilon \rho
_{s},\quad \varepsilon \in \left[ 0,1\right] .
\end{equation}%
Obviously, $\rho _{\varepsilon }\in {\cal S}$ for each $\varepsilon $. In
the geometrical view, $\rho _{\varepsilon }$ is on the line segment
connecting $\rho _{s}$ and $\rho _{0}$, Then consider another form of
mixture of $\rho _{s}$ and $\rho _{0}$,

\begin{equation}
\rho _{\varepsilon ^{^{\prime }}}=\left( 1-\varepsilon ^{\prime }\right)
\rho _{0}+\varepsilon ^{\prime }\rho _{s}\qquad \varepsilon ^{\prime }>1.
\end{equation}%
At this time, $\rho _{\varepsilon ^{\prime }}$ is on the extension line from 
$\rho _{0}$ to $\rho _{s}$. From the theorems in \cite{20}, we know that $%
\rho _{\varepsilon ^{\prime }}$ can be in the interior of ${\cal S}$, on the
boundary of ${\cal S}$, or outside of ${\cal S}$. That depends on the values
of $\varepsilon ^{\prime }$. So for sufficiently large $\varepsilon ^{\prime
}>1$, $\rho _{\varepsilon ^{\prime }}$ may be outside of ${\cal S}$.

Let us now obtain our main result, which refer to the boundary of ${\cal S}$.

\medskip

{\bf THEOREM }For a given separable state $\rho _b\in {\cal S}$, if and only
if either of the following conditions is satisfied, $\rho _b$ is on the
boundary of ${\cal S}$ ,

$\left( i\right) $ there exists at least one vanishing eigenvalue for $\rho
_b$;

$\left( ii\right) $ there exists at least one vanishing eigenvalues for $%
\widetilde{\rho }_b$.

or in the other words, the rank of $\rho _b$ or $\widetilde{\rho }_b$ is no
larger than 3.

{\it Proof } Note that ${\cal S\in H}$ and ${\cal S}$ is compact and convex
and that $\rho _{0}$ is in the interior of ${\cal S}$. If $\rho _{b}$ is on
the boundary of ${\cal S}$, let us consider the mixture with form of Eq.(8),
that is,

\begin{equation}
\rho _{\delta }=(1-\delta )\rho _{0}+\delta \rho _{b},\quad \delta >1.
\end{equation}%
From \cite{20}, we know that such $\rho _{\delta }$ does not belong to $%
{\cal S}$. Therefore, for a given ${\cal \delta }>1$, there are two
possibilities for $\rho _{\delta }$ . That is, either $\rho _{\delta }$ is
positive and thus represents an entangled state, or $\rho _{\delta }$ is
non-positive and does not represent any physical state. In the former case,
any point on the line segment from $\rho _{0}$ to $\rho _{\delta }$
represents a real physical state, and particularly, all the points (except $%
\rho _{b}$) on the line segment from $\rho _{b}$ to $\rho _{\delta }$ are
entangled states. Let $\delta \rightarrow 1$, then $\rho _{\delta
}\rightarrow \rho _{b}$. Correspondingly $\widetilde{\rho }_{\delta
}\rightarrow \widetilde{\rho }_{b}$. Under this condition, the distance
between $\rho _{b}$ and $\rho _{\delta }$ can be arbitrarily small. Note
that both $\rho _{b}$ and $\widetilde{\rho }_{b}$ are clearly positive and
that $\rho _{\delta }$ is also positive. But $\widetilde{\rho }_{\delta }$
is non-positive. So by connectivity of the space and continuity of $\delta $
we conclude that $\widetilde{\rho }_{b}$ must have vanishing eigenvalue. For
the other case in which $\rho _{\delta }$ is non-positive and does not
represent real state, similar analysis shows that $\rho _{b}$ has at least
one vanishing eigenvalue.

To prove the reverse. We first consider a separable state $\rho _{s}$ the
eigenvalues of which can be denoted by $\lambda _{1}\geq \lambda _{2}\geq
\lambda _{3}\geq \lambda _{4}=0$, that is, $\rho _{s}$ has at least one
vanishing eigenvalue. We assume that $\rho _{s}$ is in the interior of $%
{\cal S}$. Due to the compactness and convexity of ${\cal S}$ and with the
help of \cite{20}, we know that there always exists a $\varepsilon ^{\prime
}>1$ such that the mixture $\rho _{\varepsilon ^{\prime }}$ in the form $%
\rho _{\varepsilon ^{\prime }}=(1-\varepsilon ^{\prime })\rho
_{0}+\varepsilon ^{\prime }\rho _{b}\quad (\varepsilon ^{\prime }>1)$ is
also in the interior of ${\cal S}$. The eigenvalues of $\rho _{\varepsilon
^{\prime }}$, denoted by $\lambda _{i}^{\prime }$, can be easily calculated,
that is, 
\begin{equation}
\lambda _{i}^{\prime }=\varepsilon ^{\prime }\lambda _{i}+\frac{%
1-\varepsilon ^{\prime }}{4},\quad \left( i=1,\cdots ,4\right) .
\end{equation}

Obviously, for $\lambda _{4}=0$, we have $\lambda _{4}^{\prime }=\frac{%
1-\varepsilon ^{\prime }}{4}<0$. Then $\rho _{\varepsilon }^{\prime }$ is
non-positive and can not belong to ${\cal S}$. This result contradicts with
the assumption made above. Hence $\rho _{s}$ can not be in the interior and
can only be on the boundary of ${\cal S}$, i.e. $\rho _{s}=\rho _{b}$. On
the other hand, let's consider a separable state $\rho _{s}$ such that $%
\widetilde{\rho }_{s}$ has one vanishing eigenvalue, {\it i.e.} $\widetilde{%
\lambda }_{1}\geq \widetilde{\lambda }_{2}\geq \widetilde{\lambda }_{3}>%
\widetilde{\lambda }_{4}=0$. Also assume that $\rho _{s}$ is in the interior
of ${\cal S}$. Then $\rho _{\varepsilon ^{\prime }}$ is the interior point
of ${\cal S}$ for some $\varepsilon ^{\prime }>1$. So $\widetilde{\rho }%
_{\varepsilon ^{\prime }}$ is necessarily positive. Denote the eigenvalues
of $\widetilde{\rho }_{\varepsilon ^{\prime }}$ by $\widetilde{\lambda }%
_{i}^{\prime },i=1,\cdots 4$. Similar analysis will demonstrate that one of
the eigenvalues of $\widetilde{\rho }_{\varepsilon ^{\prime }}$, which is
actually $\widetilde{\lambda }_{4}$, is negative. It is contrary to the
assumption. So $\rho _{s}=\rho _{b}$.

\section{DISCUSSION}

Having obtained the description of the boundary of ${\cal S}$, we now return
to think about some concepts of quantum entanglement.

\subsection{Relative Entropy of Entanglement}

In \cite{15,16}, Vedrel {\it et.al.} propose the concept of relative entropy
of entanglement and give the numberical results for some examples of
two-qubit entangled states. For instance, given entangled state $\rho _{e}$
as%
\begin{equation}
\rho _{e}=\lambda \left| \Phi ^{+}\right\rangle \left\langle \Phi
^{+}\right| +\left( 1-\lambda \right) \left| 01\right\rangle \left\langle
01\right| ,\quad \lambda \in \left( 0,1\right) ,
\end{equation}%
where $\left| \Phi ^{+}\right\rangle =\frac{1}{\sqrt{2}}\left( \left|
00\right\rangle +\left| 11\right\rangle \right) $. The separable state which
minimize the relative entropy $S\left( \rho _{e}\parallel \rho _{s}\right) $
(cf. Eq.(1)) is said to be%
\begin{eqnarray}
\rho _{s} &=&\frac{\lambda }{2}\left( 1-\frac{\lambda }{2}\right) \left|
00\right\rangle \left\langle 00\right| +\frac{\lambda }{2}\left( 1-\frac{%
\lambda }{2}\right) \left( \left| 00\right\rangle \left\langle 11\right|
+H.C.\right)   \nonumber \\
&&+\left( 1-\frac{\lambda }{2}\right) ^{2}\left| 01\right\rangle
\left\langle 01\right| +\frac{\lambda ^{2}}{4}\left| 10\right\rangle
\left\langle 10\right| +\frac{\lambda }{2}\left( 1-\frac{\lambda }{2}\right)
\left| 11\right\rangle \left\langle 11\right| 
\end{eqnarray}%
Straightforward computation shows that $\rho _{s}$ has a vanishing
eigenvalue. Then $\rho _{s}$ is on the boundary of ${\cal S}$. This result
also holds in the other examples mentioned in \cite{15,16}. So we can say,
at least for these specific cases, the relative entropy of entanglement can
be achieved on the boundary of the set of separable states. It is possible
to generalize this result.

\subsection{Robustness of Entanglement}

Recall that the robustness of entanglement proposed by Vidal {\it et al.}%
\cite{17}. The geometrical interpretation has been obtained in \cite{21}, in
which the boundary of ${\cal S}$ is not yet explicitly defined. The
discussion in this Letter can be supplementary to \cite{21}. We can say that
for a given entangled state $\rho _{e}$ of two-qubit system, if $s$ in
Eq.(2) is minimal among all separable states $\rho _{s}$, {\it i.e. }$s$ is
the absolute robustness of $\rho _{e}$, then $\rho _{s}$ and $\rho \left(
s\right) $ in Eq.(2) are necessarily on the boundary of ${\cal S}$.

\subsection{Lewenstein-Sanpera Decomposition}

Now let us give our interpretation for L-S decomposition. Recall the form of
L-S decomposition represented by Eq.(3). Using Hilbert-Schmidt metric
defined by Eq.(6), we can express the $\lambda $ in Eq. (3) as 
\begin{equation}
\lambda =\frac{\left\| \rho -P_{e}\right\| }{\left\| \rho _{s}-P_{e}\right\| 
}.
\end{equation}%
As shown in Fig.1, $\lambda $ can be regarded as the ratio of the length of
line segment $AB$ to that of $AC$. With $P_{e}$ (point $A$ in Fig. 1) and $%
\rho $ (point $B$) fixed, $\lambda $ reaches the maximal when $\rho _{s}$ is
on the boundary of ${\cal S}$, {\it i.e.}, the length of $AC$ is minimal. In
the geometrical view, this result is straightforward. So we can always
rewrite Eq. (5) as 
\begin{equation}
\rho =\eta \rho _{b}+\left( 1-\eta \right) P_{e},\quad \lambda \in \left[ 0,1%
\right] ,
\end{equation}%
where $\rho _{b}$ is on the boundary and $\eta $ is the maximal $\lambda $
associated with $P_{e}$. Obviously,for Eq.(4) $\rho _{s}^{(opt)}$ is on the
boundary of ${\cal S}$ and can be written as $\rho _{b}^{(opt)}$ while $S$
can be expressed as $S=\max \left\{ \eta \right\} $.

In \cite{22}, Englert {\it et al.} have erred to what is involved in L-S
decomposition, that is, if $\rho =S\rho _{b}^{(opt)}+(1-S)P_{e}^{(opt)}$ is
the optimal decomposition, then (a) the state $(1+\varepsilon )^{-1}(\rho
_{b}^{(opt)}+\varepsilon P_{e}^{(opt)})$ is non-separable for $\varepsilon >0
$; and (b) the state $\rho _{b}^{(opt)}+(\frac{1}{S}%
-1)(P_{e}^{(opt)}-P_{e}^{\prime })$ is either non-positive or non-separable
for each $P_{e}^{\prime }\neq P_{e}^{(opt)}$. It is said in \cite{16} that
their verification is rather complicated even in seemingly simple cases. But
in our view, it is easy to understand. Now we give our explanation as
follows.

For conclusion (a), since the optimal decomposition of $\rho $ has been
obtained, $\rho _{b}^{(opt)}$ is necessarily on the boundary of ${\cal S}$.
Thus the convexity of ${\cal S}$ guarantees that all points on the line
segment {\it AD} (cf Fig. 1) can not belong to ${\cal S}$ and without choice
represent entangled states. These states are expressed as the convex sum of $%
P_{e}^{(opt)}$ and $\rho _{b}^{(opt)}$, i.e., $(1+\varepsilon )^{-1}(\rho
_{b}^{(opt)}+\varepsilon P_{e}^{(opt)}),\varepsilon >0$. In addition, this
conclusion also holds without the ``optimal'' condition. That is, for any
decomposition determined by Eq. (14), conclusion (a) is valid.

For conclusion (b), let 
\begin{equation}
\rho _{x}=\rho _{b}^{(opt)}+\frac{1-S}{S}(P_{e}^{(opt)}-P_{e}^{\prime }),
\end{equation}%
or equivalently, 
\begin{equation}
S\rho _{x}+(1-S)P_{e}^{^{\prime }}=S\rho _{b}^{(opt)}+(1-S)P_{e}^{(opt)}.
\end{equation}%
The right-handed of Eq. (16) is just the optimal decomposition of $\rho $.
We already know that $S$ is the largest $\eta $ of all possible
decomposition (cf Eq.(14)). If the mixture on the left-hand of Eq.(16) does
represent $\rho $, the weight of $\rho _{x}$ must be less than $S$. In other
words, the left-hand of Eq.(16) can not be a real L-S decomposition about $%
\rho $. Hence $\rho _{x}\notin {\cal S}$. Theore, either $\rho _{x}$ can not
be positive, or can represent a non-separable state. The conclusion (b) can
be demonstrated in Fig. 2. Of course the ``optimal'' condition is necessary
in this case.\medskip 

So far we have discussed the boundary of the set of all separable states.
Using this concept, we give straightforward explanations for some
entanglement measure of the mixed states, mainly for L-S decomposition. We
feel that introducing boundary may be helpful for better understanding of
entanglement.

\begin{acknowledgement}
This project is supported by the National Nature Science Foundation of China
(10075041 and 10075044) and the Science Foundation of USTC for Young
Scientists.
\end{acknowledgement}

\bigskip

Figure Caption

Fig. 1 $P_e$ (point $A$) denotes an entangled pure state. $\rho $ (point $B$%
) represents a given non-separable mixed state. ${\cal S}$ is the set of all
separable states. $\rho _s$ (point $C$) is in the interior of ${\cal S}$ and 
$\rho _b$ (point $D$) is on the boundary of ${\cal S}$.

Fig. 2 $P_{e}^{(opt)}$ and $P_{e}^{\prime }$ are two different entangled
pure states. $P_{e}^{(opt)}$ and $\rho _{b}^{(opt)}$ construct the optimal
L-S decomposition of $\rho $. $\rho _{x}$ is outside of set ${\cal S}$. $%
P_{e}^{\prime }$ and $\rho _{x}$ does not construct the L-S decomposition of 
$\rho $.

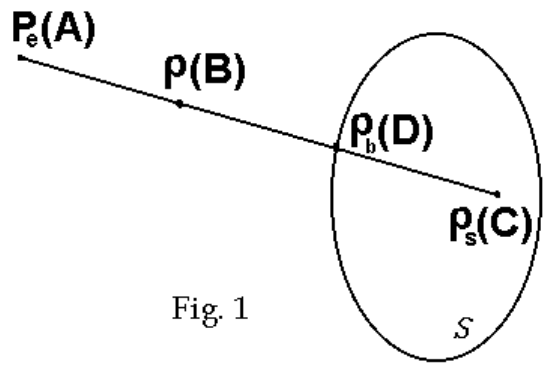
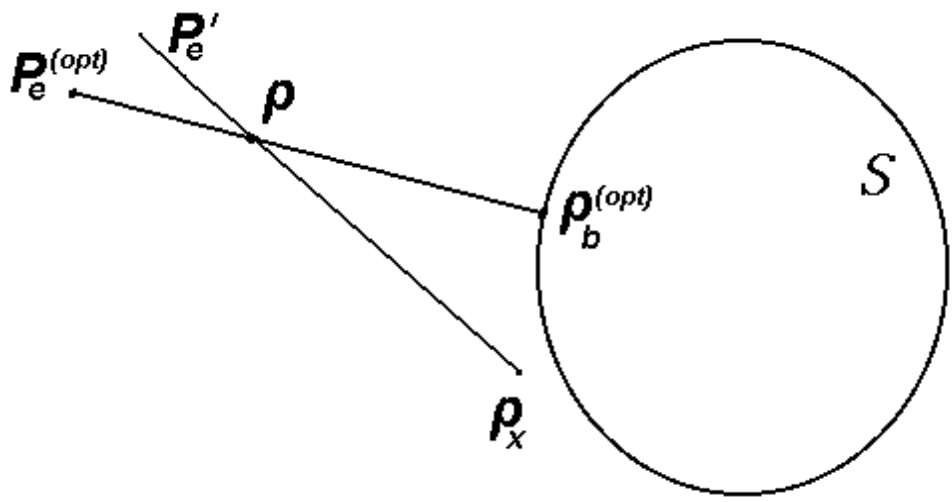
\end{document}